\documentclass[review]{elsarticle}

\usepackage{hyperref}

\newproof{pf}{Proof}
\newdefinition{rmk}{Definition}

\usepackage{amsmath}
\usepackage{amssymb}
\DeclareSymbolFontAlphabet{\mathcal}   {symbols}
\usepackage{natbib}
\usepackage{graphicx}
\usepackage{algorithm}
\usepackage[noend]{algpseudocode}
\usepackage{caption}
\usepackage{subcaption}
\usepackage{multirow}
\usepackage{hhline}
\usepackage[T1]{fontenc}
\usepackage[utf8]{inputenc}
\usepackage[numbers]{natbib}

\DeclareMathOperator*{\argmax}{arg\,max}

\journal{Journal of Information Sciences}

\begin{document}

\begin{frontmatter}

\title{Asymptotic resolution bounds of generalized modularity and multi-scale community detection}

\author[1,2]{Xiaoyan Lu}
\author[1,2]{Brendan Cross}
\author[1,2,3]{Boleslaw K. Szymanski\corref{mycorrespondingauthor}}
\cortext[mycorrespondingauthor]{Corresponding author}
\ead{szymab@rpi.edu}

\address[1]{Department of Computer Science, Rensselaer Polytechnic Institute, Troy, NY, USA}
\address[2]{Network Science and Technology Center, Rensselaer Polytechnic Institute, Troy, NY, USA}
\address[3]{Spo\l{}eczna Akademia Nauk, \L{}\'{o}d\'{z}, Poland}

\begin{abstract}
The maximization of generalized modularity performs well on networks in which the members of all communities are statistically indistinguishable from each other. However, there is no theory bounding the maximization performance in more realistic networks where edges are heterogeneously distributed within and between communities. Using the random graph properties, we establish asymptotic theoretical bounds on the resolution parameter for which the generalized modularity maximization performs well. From this new perspective on random graph model, we find the resolution limit of modularity maximization can be explained in a surprisingly simple and straightforward way. Given a network produced by the stochastic block models, the communities for which the resolution parameter is larger than their {\it densities} are likely to be spread among multiple clusters, while communities for which the resolution parameter is smaller than their background inter-community edge {\it density} will be merged into one large component. Therefore, no suitable resolution parameter exits when the intra-community edge {\it density} in a subgraph is lower than the inter-community edge {\it density} in some other subgraph. For such networks, we propose a progressive agglomerative heuristic algorithm to detect practically significant communities at multiple scales.
\end{abstract}


\end{frontmatter}

\section{Introduction}
In complex networks, community structures are widely observed. Detecting such community structures can be viewed as partitioning of the network into clusters in which the nodes are more densely connected to each other than to the nodes in the rest of the network. Modularity maximization~\cite{newman2006modularity} is one of the state-of-the-art methods for community detection. It aims at discovering the partition of the network that maximizes modularity, a well-known quality measure of network community structure.

Modularity maximization, however, suffers from the so-called resolution limit problem~\cite{fortunato2007resolution,lancichinetti2011limits} that is this method's tendency to detect communities of similar properties. In such cases, standard modularity reaches maximum by combining some small well-formed communities into inappropriate large clusters or by spreading some large well-formed communities among smaller ones. Some variants of the modularity function have been proposed either to resolve this problem~\cite{chen2014community,lu2018adaptive,traag2011narrow} or to enable detection of communities at different scales~\cite{lewis2010function,simon1991architecture,porter2009communities,traag2013significant}. A popular choice for the latter is the {\em generalized modularity} of Reichardt and Bornholdt~\cite{reichardt2006statistical}, which scales the discovered community sizes according to a simple resolution parameter.

Besides the maximization of modularity and its generalized version, an alternative approach to detect communities is the statistical inference to fit the generative model to the observed network data. The inference assumes that the random graph model produced the observed network provided as input. The statistical inference aims at recovering a partition that maximizes the likelihood of the random graph model generating the observed network data.

\begin{figure}[!ht]
    \centering
    \includegraphics[width=9cm]{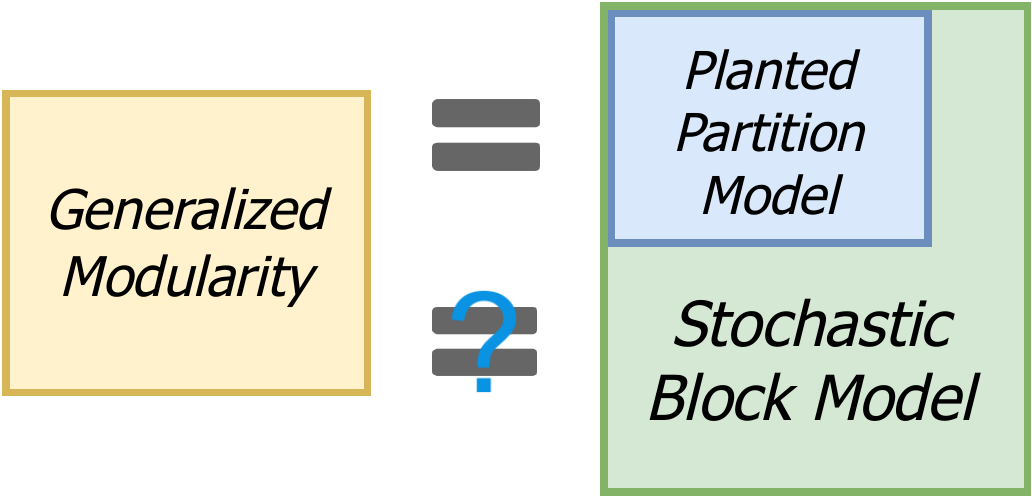}
    \caption{The maximum-likelihood estimation of the degree-corrected planted partition model and the maximization of generalized modularity with the corresponding resolution parameter recover the equivalent communities~\cite{newman2016equivalence}. Our paper answers the important question about the performance of the generalized modularity maximization on the networks generated by the degree-corrected stochastic block model.}
    \label{fig:idea}
\end{figure}

One widely used generative model for community structure is the degree-corrected stochastic block model~\cite{karrer2011stochastic} where nodes are organized into blocks while edges are placed between the nodes independently at random\footnote{In the Supplementary Material, we discuss in detail various stochastic block model variants and the related work on modularity maximization in the literature.}. The maximization of the generalized modularity~\cite{newman2016equivalence} is shown to be equivalent to the maximum-likelihood estimation (MLE) of the degree-corrected planted partition model, a special case of the degree-corrected stochastic block model, on the same graph. Yet, there is no asymptotic bounds defining generalized modularity maximization's performance in more realistic networks generated by the degree-corrected stochastic block model. As illustrated by Fig.~\ref{fig:idea}, our work aims at answering the important question about the performance of the generalized modularity on such networks.



Specifically, we establish here an asymptotic theoretical upper and lower bounds on the resolution parameter of generalized modularity. This result bridges the gap between the literature on the resolutions limits of modularity-based community detection~\cite{fortunato2010community,fortunato2007resolution} and the random graph models~\cite{karrer2011stochastic,newman2016equivalence}. Given a resolution parameter within the established range, we show that maximizing the generalized modularity is still likely to detect the correct communities, regardless whether the equivalence between generalized modularity maximization and the MLE of stochastic block model holds or not. When the resolution parameter of the generalized modularity is larger than the upper bound we developed, some well-formed communities are likely to be spread among multiple clusters to increase the generalized modularity. However,  when the resolution parameter of generalized modularity is smaller than the lower bound we developed, some communities may be inappropriately merged into one large component.

The experimental results of modularity maximization on synthetic network agree with our findings that the generalized modularity performs best when resolution parameter lies well within the interval defined by our derived analytical bounds. Experimentally, and without any explanation why, the authors of~\cite{fenn2009dynamic,mucha2010community,traag2013significant} established that the suitable values of the resolution parameter occur in the widest interval in which the generalized modularity maximization produces the consistent partitions. 
The authors of~\cite{traag2011narrow} introduce the class of quality measures that are "resolution-limit-free". The generalized modularity does not fall into this category because the resolution parameter performs best within an interval conditioned on the number of clusters on the ring network of cliques~\cite{fortunato2007resolution}. Our work derives such interval of the resolution parameter in the boarder degree-corrected stochastic block graphs, 
thus connecting the literature on resolutions bounds for community detection~\cite{fortunato2007resolution,fortunato2010community}
with the multi-scale community discovery~\cite{fenn2009dynamic,mucha2010community,young2015shadowing}.

Furthermore, our findings shed light on the adaptation of the generalized modularity to networks in which there is no universal resolution parameter to detect all communities. That happens only when the lower bound of the resolution parameter may be higher than its upper bound. Therefore, some well-formed communities either are split into multiple clusters or merged into one large component. The problem is analogous to identifying mountains with varied heights of their peaks and valleys by scanning mountainsides at a single altitude. Such scan either would miss the low peaks seeing just background behind them, or would treat the higher peaks as one mountain as their mountainsides would form single base unbroken by valleys.

To address the above-mentioned problem, we propose a progressive agglomerative heuristic algorithm that systematically adjusts the resolution parameter. The algorithm recursively splits the resulting clusters of the previous level to detect smaller communities. As the recursion proceeds, the algorithm gradually increases the resolution parameter for high-resolution community detection in local clusters of the network. The algorithm proceeds until none of the currently found clusters contains communities that are practically significant as measured by Bayes model selection~\cite{chipman2001practical}. Compared to the algorithms using a uniform resolution parameter, our approach does not require multiple re-estimation of the resolution parameter for the entire network~\cite{newman2016equivalence} that can be computationally prohibitively costly for large networks.

To summarize, this paper makes the following contributions to our understanding of multi-scale community detection, resolution limit and generalized modularity:
\begin{itemize}
    \item Asymptotic upper and lower bounds on the resolution parameter of generalized modularity linking the resolutions limits of modularity-based community detection~\cite{fortunato2010community,fortunato2007resolution} to the random graph models~\cite{karrer2011stochastic,newman2016equivalence}.
    \item A simple theoretical explanation of the resolution limit, confirming the previous experimental and theoretical findings~\cite{fenn2009dynamic,mucha2010community,traag2013significant,traag2013significant} that the suitable resolution parameters occur in the widest interval in which the generalized modularity maximization produces the consistent partitions.
    \item A highly efficient multi-scale modularity-based heuristic community detection algorithm that does not require multiple re-estimation of the resolution parameter.
\end{itemize}

\section{Approach} \label{sec:approach}

\subsection{Asymptotic bounds on resolution parameter}
The generalized modularity of Reichardt and Bornholdt~\cite{reichardt2006statistical} can be defined as
\begin{equation} \label{eq:modularity_definition_alternative}
    Q(\gamma) = \sum_{r} \left[\frac{m_r}{m} - \gamma \left(\frac{ \kappa_r }{ 2m } \right)^2 \right],
\end{equation}
where $m_r$ is the number of edges with both endpoints inside the community $r$, $\kappa_r$ is the sum of the degrees of nodes in community $r$, and $m$ is the total number of edges in the network.

Merging two different communities, $r$ and $s$, results in the following equations: the total number of edges inside the merged community becomes $m_{r\cup s}=m_r+m_s+m_{r,s}$ where $m_{r,s}$ is the number of edges between communities $r$ and $s$. The sum of degrees of the nodes inside the merged community is $\kappa_{r\cup s} = \kappa_r+\kappa_s$. Hence, given the formalization of the generalized modularity in Eq.~(\ref{eq:modularity_definition_alternative}), the optimization algorithm is able to detect two well-formed communities $r$ and $s$ if the change of generalized modularity from merging $r$ and $s$ is non-positive, leading to the inequality
\begin{equation}
\Delta Q(\gamma) = \frac{m_{rs}}{m} - \gamma \frac{\kappa_{r}\kappa_{s}}{2m^2} \leq 0,
\end{equation}
which can be rewritten in the alternative way as
\begin{equation}\label{eq:resolution_0}
\kappa_{r}\kappa_{s} 
\geq 2 \frac{m_{rs} m} {\gamma}.
\end{equation}
Otherwise, when the $\Delta Q(\gamma) > 0$, communities $r$ and $s$ are merged to increase $Q(\gamma)$. Clearly, one can always increase $\gamma$ so that Eq.~(\ref{eq:resolution_0}) holds for any small $\kappa_r$ and $\kappa_s$. Howether, a large $\gamma$ may result in inappropriate split of some communities. To see this point, consider a community $t$ comprised of two sets of nodes $t'$ and $t''=t-t'$ with sums of degrees $\kappa'$ and $\kappa''$ respectively, and $m''$ edges between nodes in $t'$ and $t''$. To avoid splitting community $t$ into $t'$ and $t''$, the inequality
\begin{equation}~\label{eq:resolution_1}
\kappa'\kappa'' \leq 2\frac{m''m}{\gamma}
\end{equation}
must hold. Given Eq.~(\ref{eq:resolution_0}) and Eq.~(\ref{eq:resolution_1}), we have
\begin{equation} ~\label{eq:resolution}
  \frac{\kappa'\kappa''}{m''} \leq \frac{2m}{\gamma} \leq \frac{\kappa_r \kappa_s}{m_{rs}}.  
\end{equation}
A simple and straightforward explanation of the resolution limit found by Fortunato et al.~\cite{fortunato2007resolution} is that in realistic networks the above inequality may not hold. Indeed, it is likely that in a large network there exist communities $r$ and $s$ with small $\kappa_r$ and $\kappa_s$ and community $t$ with large sum of node degrees $(\kappa'+\kappa'')$, such that the inequality above does not hold because $\frac{\kappa'\kappa''}{m''} > \frac{\kappa_r \kappa_s}{m_{rs}}$, giving rise to resolution limit anomaly.

The degree-corrected stochastic block model\footnote{See the Supplementary Material for the formal definition of the model.} is a generative model of the graph in which nodes are organized as blocks and edges are placed between nodes independently at random. Following the notations in~\cite{newman2016equivalence}, the degree-corrected stochastic block model assumes the number of edges between nodes $i$ and $j$ follows the Poisson distribution with the mean defined as 
\begin{equation} \label{eq:sbm_newman_definition}
\eta_{ij} = \omega_{g_i g_j} \frac{k_i k_j}{2m},
\end{equation}
where for node $l$, $g_l$ is the block assignment of this node and $k_l$ is its degree. Note that the model defines $\eta_{ii} = \omega_{g_i g_i} \frac{k_i k_i}{4m}$, i.e., a half the number for the self-edge. Given the nodes' degrees $k_i$ and $k_j$, $\frac{k_i k_j}{2m}$ is the expected number of edges between different nodes $i$ and $j$, or a half that number for self-edges, in the graph ensembles generated by the configuration model~\cite{molloy1995critical}. Thus, $\omega_{rs}$ represents the ratios of the expected numbers of edges in the stochastic block model and the configuration model. In the rest of the paper, we refer to the parameter $\omega_{rs}$ as the edge {\it density} between blocks $r$ and $s$ and to the matrix ${\bf \Omega} = \{\omega_{rs}\}$ as the {\it density} matrix of the model.

The expected number of edges between two communities $r$ and $s$ in the degree-corrected stochastic block model is
\begin{equation} \label{eq:approx_out2}
    m_{rs} \approx \sum_{\{i,j\}\in V_r \times V_s} \eta_{ij} = \sum_{\{i,j\}\in V_r \times V_s} \omega_{rs} \frac{k_i k_j}{2m} = \omega_{rs} \frac{\kappa_r \kappa_s}{2m}.
\end{equation}

The expected number of edges between two subsets of nodes $t'$ and $t''=t - t'$ inside the same community $t$ is
\begin{equation} \label{eq:approx_in2}
    m_{t't''} \approx \sum_{\{i,j\}\in t'\times t''} \eta_{ij} = \sum_{\{i,j\}\in t'\times t''} \omega_{tt} \frac{k_i k_j}{2m} = \omega_{tt} \frac{\kappa_{t'} \kappa_{t''}}{2m},
\end{equation}
where $\omega_{tt}$ is the $t^{th}$ diagonal element in the {\it density} matrix.

Using the resolution inequality from Eq.~(\ref{eq:resolution}), these approximations lead to the range
\begin{equation} \label{eq:sbm_bounds}
    \max_{r\neq s} \omega_{rs} \leq \gamma \leq \min_{t} \omega_{tt}
\end{equation}
within which a uniform $\gamma$ value avoids the resolution limit trap. This result connects the resolution limit of generalized modularity with the random graph models.

Eq.~(\ref{eq:sbm_bounds}) indicates that the suitable $\gamma$ value should be as small as possible to avoid splitting of any well-defined communities, i.e. $\gamma \leq \omega_{tt}$ for any $t$. Otherwise, when $\gamma$ is larger than the {\it density} parameter of some loose community $t$, this community $t$ is likely to be split. However, $\gamma$ should not be larger than any background inter-community {\it density}, i.e. $\omega_{rs} \leq \gamma$ for every pair of $r \neq s$. Otherwise, the communities $r$ and $s$ are likely to be merged inappropriately into one cluster to maximize the generalized modularity.

The degree-corrected planted partition model is a special case of the degree-corrected stochastic block model where $\omega_{rs} = \omega_{out}$ for every pair of different communities $r\neq s$ and $\omega_{tt} = \omega_{in}$ for every community $t$. Therefore, Eq.~(\ref{eq:sbm_bounds}) also applies to the graphs generated by the degree-corrected planted partition model. In such a case, Eq.~(\ref{eq:sbm_bounds}) becomes $\omega_{out} \leq \gamma \leq \omega_{in}$. As shown in ~\cite{newman2016equivalence}, when $\gamma = \frac{\omega_{in} - \omega_{out}}{\log \omega_{in} - \log \omega_{out}}$, the generalized modularity maximization is equivalent to the maximum likelihood estimation of the degree-corrected planted partition model. Using this equivalence condition in the inequality, we obtain  
\begin{equation} \label{eq:ppm_bounds_est}
    \omega_{out} \leq \frac{\omega_{in} - \omega_{out}}{\log \omega_{in} - \log \omega_{out}} \leq \omega_{in}
\end{equation}
which can be proven\footnote{Dividing Eq.~(\ref{eq:ppm_bounds_est}) by $\omega_{out} > 0$ on both sides and denoting $x=\omega_{in}/\omega_{out} > 1$ leads to inequality $1 \leq (x - 1)/\log(x) \leq x$ which can be formally proved using the well-known upper and lower bounds of the natural logarithm $1 - 1/x\leq \log x \leq x - 1$ for any $x > 1$.} given any positive $\omega_{out} < \omega_{in}$ values. Hence, the maximum likelihood estimation of the degree-corrected planted partition model always resides within the bounds we derive here.

\subsection{Plateaus problem}
In a network generated by the degree-corrected stochastic block model with $\omega_{rs} > \omega_{tt}$ for some $r \neq s$ and $t$, Eq.~(\ref{eq:sbm_bounds}) indicates that a uniform resolution parameter is not sufficient for the recovery of communities $r$, $s$ and $t$. Given the ground truth communities, the posterior estimates of the {\it density} matrix $\Omega$ is $\hat{\omega}_{rs} = \frac{2 m_{rs} m}{\kappa_r \kappa_s}$ for each pair of communities $r\neq s$ or $\hat{\omega}_{rr} = \frac{4 m_{r} m}{\kappa_r^2}$ for each community $r$. To ensure the correct recovery of the ground truth, the resolution parameter should satisfy
\begin{equation} \label{eq:sbm_bound_explicity}
    \frac{2 m_{rs} m}{\kappa_r \kappa_s} \leq \gamma \leq \frac{4 m_{l} m}{\kappa_l^2}
\end{equation}
for any communities $r\neq s$ and $l$. 

The classical example of resolution limit trap is presented in~\cite{lancichinetti2011limits} where an undirected unweighted network contains three communities: two cliques and one random graph, and every two communities are connected by one single edge. Suppose each clique includes 6 nodes and the random graph contains 100 nodes and 956 edges. Given the three communities, the posterior estimation of the {\it density} matrix $\Omega$ of the degree-corrected stochastic block model is
\begin{equation}
    \Omega = \begin{bmatrix}
    1.03	& 0.03  &	0.03	\\
    0.03	& 57.94 &	1.93  \\
    0.03	& 1.93  &	57.94 \\
    \end{bmatrix},
\end{equation}
where the first row and column correspond to the random graph while the remaining rows and columns correspond to the two cliques respectively. There is no suitable resolution parameter $\gamma$ to detect three communities in this case because the {\it density} parameter for the edges between two cliques $1.93$ is larger than the {\it density} parameter for the edges inside random graph $1.03$. When applying generalized modularity maximization, adopting a resolution parameter larger than $1.93$, makes it likely that two cliques will be detected, but the random graph will get split into smaller communities. On the other hand, a resolution parameter within $[1.03, 1.93]$ preserves the random graph as one complete community, but the two cliques will be merged into one community.

\begin{figure}
    \centering
    \includegraphics[width=0.9\textwidth]{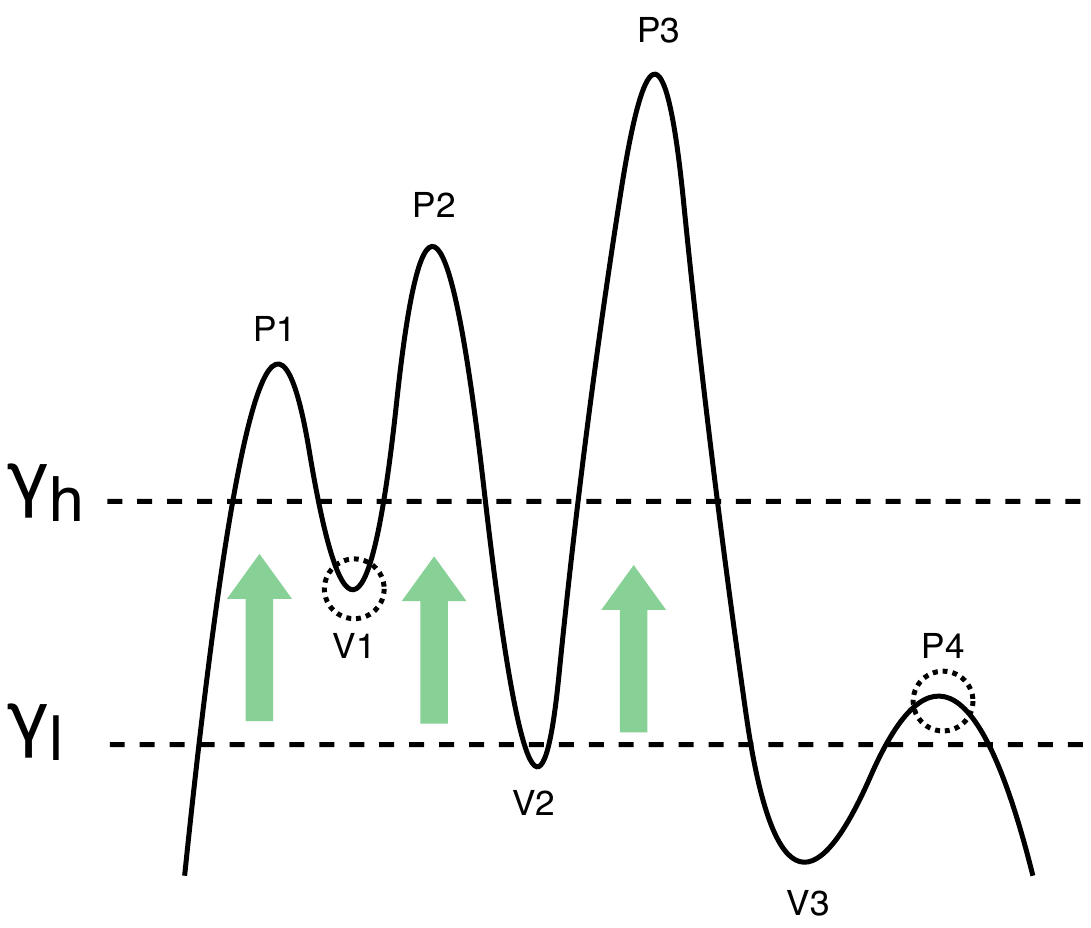}
    \caption{The plateaus problem is analogous to searching for mountains that are located at different plateaus. Using a single altitude either would miss the lower mountains, or would treat the higher peaks as one mountain. Specifically, when using resolution parameter $\gamma_l$, the left two high peaks P1 and P2 are considered one ``mountain" - two well-formed dense communities are merged because their inter-community edge density (illustrated by valley V1) is higher than $\gamma_l$. If we adopt a higher resolution parameter $\gamma_h$, the low peak P4 on the right is ignored - a loose community is split into multiple smaller clusters. Notably, this issue cannot be avoided as long as the valley V1 of the left two peaks P1, P2 is higher than the height of the right-most peak P4.}
    \label{fig:plateaus}
\end{figure}

Intuitively, the issue is analogous to searching for mountains that are located at different plateaus while using a single altitude, see Fig.~\ref{fig:plateaus}. Such search either would miss the lower mountains, or would treat the higher peaks as one mountain. Specifically, when using resolution parameter $\gamma_l$, the left two high peaks in Fig.~\ref{fig:plateaus} are considered one ``mountain" - two well-formed dense communities are merged. If we adopt a resolution parameter $\gamma_h$, the low peak on the right is ignored - a loose community is split into multiple smaller clusters. Notably, this issue cannot be avoided as long as the valley of the left two peaks is higher than the height of the right-most peak. 

\begin{figure}
    \centering
    \includegraphics[width=.9\textwidth]{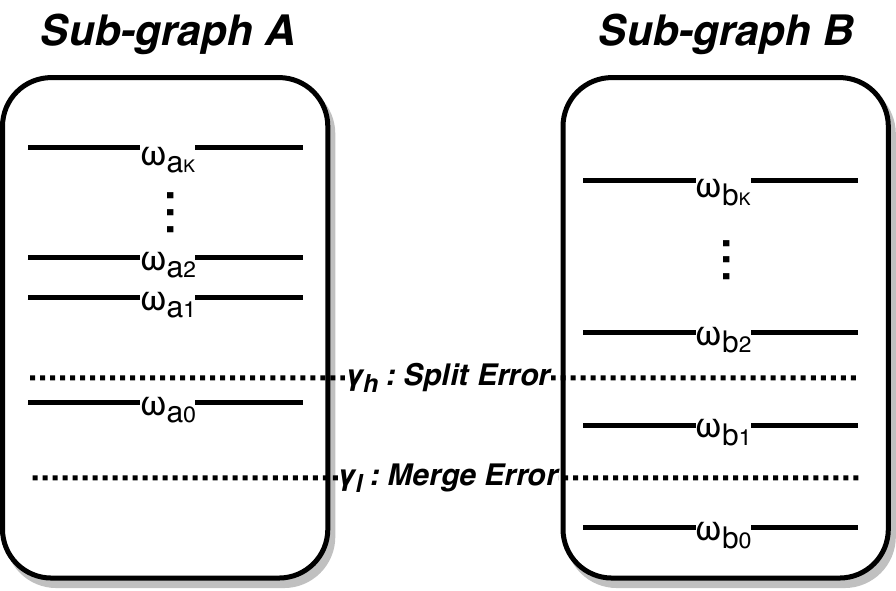}
    \caption{Resolution limits of the generalized modularity can be explained by the relations between the values of the {\it density} parameters of the degree-corrected stochastic block models. Given two disjoint subgraphs A and B such that the inter-community edge density $\omega_{a0}$ in subgraph A is larger than the intra-community edge density of some community in subgraph B, no suitable resolution parameter $\gamma$ exists because Split Error and Merge Error cannot be resolved at the same time. Split Error occurs when the resolution parameter $\gamma_h$ is larger than the inter-community edge {\it density} of a subgraph A, because the community $b_1$ with the intra-community density $\omega_{b1}$ smaller than $\gamma_h$ will be spread among multiple clusters. Merge Error occurs when the resolution parameter $\gamma_l$ is smaller than $\omega_{a0}$ so the communities in subgraph A will be merged into one community.}
    \label{fig:gamma_error}
\end{figure}
More formally, given the {\it density} matrix of a degree-corrected stochastic block model and a set of communities $S=\{r\}$, the sub-matrix $\Omega_{S,S}$ formed by the rows and columns in $r\in S$ corresponds to a subgraph in the network. Let $\omega_{w_{a0}}$ and $\omega_{w_{b0}}$ denote the inter-communities {\it density} parameter for subgraphs A and B, respectively. In Fig.~\ref{fig:gamma_error}, using $\gamma_h$ causes {\it Split Error} which splits some community with $\omega_{b1} < \gamma_h$ in  B while using $\gamma_l$ causes {\it Merge Error} which merges all communities in subgraph A.

This problem is more common in large networks than in small ones, as large networks are more likely to have inhomogeneous subgraphs. For this reason, a uniform resolution limit parameter is not sufficient to resolve communities located at different ``plateaus''. Motivated by this ``plateaus'' phenomenon, we propose a multi-scale community detection algorithm which gradually increases the resolution parameter to detect community in local subgraphs.

\subsection{\label{sec:multiscale}Multi-scale community detection}

So far, we presented the bounds on the resolution parameter for community detection in graphs generated by the degree-corrected stochastic block model. The bounds defined in Eq.~(\ref{eq:sbm_bounds}) give rise to the ``plateaus" problem when no single resolution parameter exists. When the resolution parameter takes an inappropriate value, maximizing the generalized modularity cannot discover all communities.

For this reason, we propose a heuristic algorithm that iteratively applies growing resolution parameters to detect recursively communities at different scales. This process is illustrated in Fig.~\ref{fig:plateaus}. In the first step, the generalized modularity maximization algorithm uses a small resolution parameter $\gamma_l$, which is likely to merge adjacent dense communities into clusters. Then, it applies a higher $\gamma_h$ to detect communities in each of the resulting clusters.

Specifically, at each level of recursion, the algorithm applies a small resolution parameter $\gamma_0 < 1$ in attempt to avoid inappropriately splitting of loose communities. However, it is likely to merge inappropriately
small well-formed dense communities into a large cluster. Therefore, the currently found subgraphs are passed to the next level of recursion to detect communities in them. This idea is illustrated in Fig.~\ref{fig:plateaus} where the peaks located at higher plateaus need scans at high altitude (large $\gamma_h$'s) for each to have its own community.

It is worth noting that the interval~[\ref{eq:sbm_bound_explicity}] scales with the number of edges in the graph. As the recursion proceeds to obtain smaller subgraphs, applying a resolution parameter $\gamma_{sub}$ in a subgraph with $m_{sub}$ edges is equivalent to applying $\gamma = \gamma_{sub} \frac{m}{m_{sub}} > \gamma_{sub}$ in the original graph with $m$ edges. For the simplicity, we always use the same small resolution parameter $\gamma_0 < 1$ in all subgraphs at each level of recursion. This approach is approximately equivalent to increasing the resolution parameter at each level recursion. 

The remaining challenge is to determine when to terminate the recursion. As the network breaks into smaller subgraphs recursively, the algorithm should stop when there is actually only one community in each subgraph. Indeed, one can always increase the resolution parameter to detect higher resolution communities in this subgraph. However,it does not mean the current subgraph always contains community structures. For instance, an Erdos-Renyi random graph~\cite{erds1960evolution} can be partitioned into communities as long as the resolution parameter is high enough. However, we cannot claim that such Erdos-Renyi random graph contains a community structure.

To ensure the detected communities are meaningful, we use the Bayes model selection~\cite{chipman2001practical} to evaluate the practical significance of the partitions at each level of recursion. Specifically, given a subgraph obtained at a certain recursion level, we are interested in whether or not this subgraph is more likely to be generated by the configuration model $\mathcal{H}_0$ than by a degree-corrected planted partition model $\mathcal{H}_1$ with the multiple blocks. If true, the agglomerative heuristic algorithm terminates at this recursion level and accepts the current subgraph as a single community in the result. Otherwise, the algorithm continues by applying this model selection process for each of the detected blocks.

For a particular subgraph with adjacent matrix ${\bf A}$, the null hypotheses here is that the subgraph is generated by the random graph model $\mathcal{H}_0$, which is a special case of the degree-corrected planted partition model, in which single community exists. The alternative, nested hypothesis is the degree-corrected planted partition model $\mathcal{H}_1$ along with the block assignment ${\bf g}$ found by maximization of generalized modularity.
We choose the planted partition model rather than the stochastic degree model as the alternative hypothesis because this model is generally simpler, thus leading to a Bayes posterior odds which can be efficiently computed. 
The Bayes posterior odds~\cite{chipman2001practical} can be represented as
\begin{align} \label{eq:appendix_important}
    \Lambda = \frac{P({\bf g}, \mathcal{H}_1|{\bf A})}{P(\mathcal{H}_0|{\bf A})} = 
    \frac{P({\bf A}|{\bf k,g},\mathcal{H}_1)}
    {P({\bf A}|{\bf k},\mathcal{H}_0)}
    \times
    \frac{ P({\bf g})P({\bf k})P(\mathcal{H}_1)}
    { P({\bf k})P(\mathcal{H}_0)}
\end{align}
where $P({\bf A}|{\bf k,g},\mathcal{H}_1)$ is the marginal likelihood of generating the subgraph ${\bf A}$ by $\mathcal{H}_1$, given the degree sequence ${\bf k}$, and block assignment ${\bf g}$. Likewise, $P(\mathcal{H}_0|{\bf A})$ is the marginal likelihood for the random graph model $\mathcal{H}_0$. We formulate the priors $P({\bf g}), P(\mathcal{H}_1)$ and $P(\mathcal{H}_0)$ in the Supplementary Material. Function $P({\bf k})$ does not need to be fully defined because its two appearances cancel themselves out. The logarithm of the Bayes posterior odds has a simple form of
\begin{align} \label{eq:odds}
    \ln \Lambda \approx \left[ a \ln \frac{a}{b} + (2m - a) \ln \frac{2m - a}{2m - b} \right] - N \left[ H({\bf n}) + H(B)\right]
\end{align}
where $B$ is the number communities, $n_i$ is the number of nodes in $i^{th}$ community, the entropy functions are $H({\bf n}) = - \sum_{i=1}^{B} \frac{n_i}{N} \ln \frac{n_i}{N}$ and $H(B) = - \frac{B}{N} \ln \frac{B}{N} - \frac{N - B}{N} \ln \frac{N - B}{N}$, and the coefficients $a$ and $b$ are defined as
\begin{align} 
    a &= 2\sum_r m_r  \quad \quad b = \sum_r \kappa_r^2 / 2m
\end{align}
The Supplementary Material details the mathematical derivation of $\ln \Lambda$.

Another widely used approach for model selection is the likelihood ratio test. Although the likelihood ratio test statistic is easy to obtain for $\mathcal{H}_0$ and $\mathcal{H}_1$, the null distribution of the test statistic in case of generative models used here does not follow the chi-squared distribution~\cite{yan2014model}. Thus, it requires enumerating a series of the null networks generated by $\mathcal{H}_0$ to calculate the p-value for this test, which is computationally inefficient for large networks. Therefore, we adopt the Bayes model selection to determine the termination condition at each level of the recursion.

The pseudo code of the proposed multi-scale community detection algorithm is illustrated in Alg. 1. Given a graph $G$, the algorithm maximizes the generalized modularity with a small resolution parameter $\gamma_0 < 1$, which results in a partition of the network ${\bf g}$ (Line 2). For each detected community, the algorithm tests its significance using Eq.~(\ref{eq:odds}). If such community is significant, i.e. $\ln \Lambda \leq 0$, then it is added to the result $R$. Otherwise, the multi-scale algorithm repeats the above procedure on the subgraph with nodes in this detected community. 

\begin{algorithm} 
\caption{Multi-scale Community Detection Algorithm}
\begin{algorithmic}[1] \label{algo:1}
\Procedure{multiscale\_algorithm}{$G, \gamma_0$}
\State ${\bf g} := \argmax_{g} {Q(\gamma_0)}$
\State Initialize $Res$ as an empty list

\For {each community $c$ in ${\bf g}$}
\State $\ln \Lambda := \left[ a \ln \frac{a}{b} + (2m - a) \ln \frac{2m - a}{2m - b} \right] - N \left[H({\bf n}) + H(B)\right]$
\If {$\ln \Lambda \leq 0$}
\State $Res.append(c)$
\Else
\State $G_c:=$ the subgraph of $G$ with nodes in $c$
\State ${\bf t}:=$ \textproc{multiscale\_algorithm}($G_c, \gamma_0$)
\For {each community $c$ in ${\bf t}$} 
\State $Res.append(c)$
\EndFor
\EndIf
\EndFor
\State Return $Res$
\EndProcedure
\end{algorithmic}
\end{algorithm}

\section{Experimental Results}\label{sec:experiments}

To evaluate the performance of the proposed multi-scale community detection algorithm, we compare it with the state-of-art Louvain algorithm~\cite{blondel2008fast} and CPM algorithm~\cite{traag2011narrow}, on several real and synthetic networks summarized in Table~\ref{table:summary}. 

We adopted the Louvain and CPM algorithm implementations by Vincent Traag built on top of the igraph library\footnote{https://github.com/vtraag/louvain-igraph/}. For the networks with pre-defined ground truth communities, the quality of the detected communities is evaluated using the Normalized Mutual Information (NMI) and Adjusted Rand Index (ARI) metrics that are defined in the Supplementary Material.

\begin{table}[!t]
\caption{Summary of the networks}
\label{table:summary}
\centering
\setlength\tabcolsep{2pt}
\begin{tabular}{ c|c|c|c|c|c }
 \hline
No. & Network & \#Nodes & \#Edges & Type & Ref.\\
 \hline
1 & LFR benchmark & 5K-15K & 23K-69K  & Synthetic &
~\cite{lancichinetti2008benchmark}\\
2 & Amazon product co-purchasing & 334,863 & 925,872 & Real & ~\cite{yang2015defining}\\
3 & DBLP collaboration network & 317,080 & 1,049,866 & Real & ~\cite{yang2015defining}\\
 \hline
\end{tabular}
\end{table}

\subsection{Validation of resolution bounds}
We validate the asymptotic bounds on the resolution parameters on a simplified version of the degree-corrected stochastic block model assuming the off-diagonal elements of the {\it density} matrix $\Omega$ all have the same value $\omega_0$, while the diagonal elements have different values $\omega_{rr} = \omega_r$ for each $r$. The formal definition of this so-called extended degree-corrected planted partition model can be found in the Supplementary Material.

Suppose there are $T$ communities with the {\it density} parameters monotonically ordered by their indices, according to the bounds in Eq.~(\ref{eq:sbm_bounds}), the value of the resolution parameter $\gamma$ should satisfy
\begin{equation} \label{eq:whereisgamma}
    \omega_0 \leq \gamma \leq \omega_1 < \omega_2 < .. < \omega_T,
\end{equation}
so that all $T$ communities can be detected.

\begin{figure}
    \centering
    \includegraphics[width=0.8\textwidth]{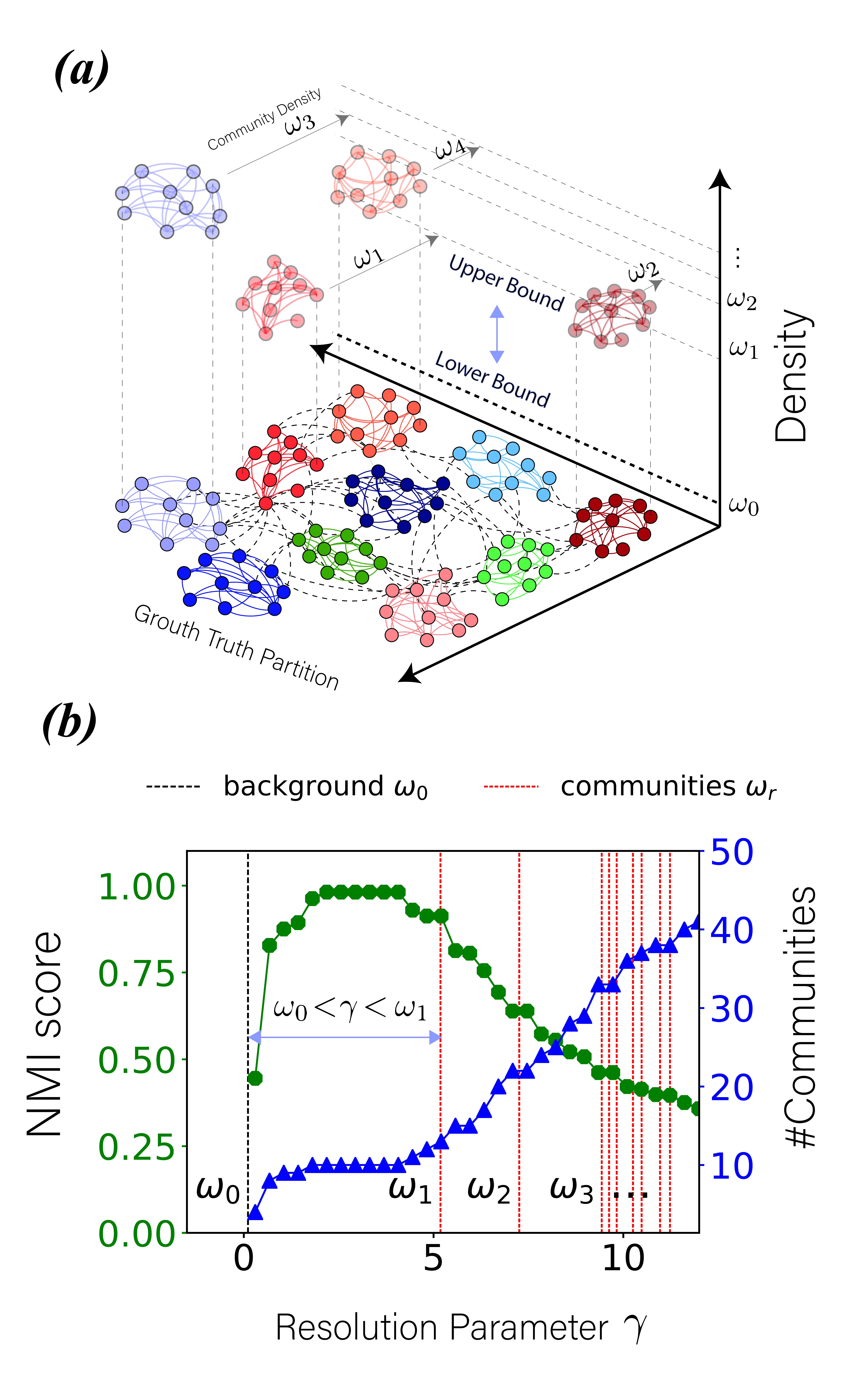}
    \caption{The generalized modularity performs well with resolution parameter in the interval $\gamma\in[1.8,4.2]$, matching the derived theoretical bound $\omega_0 \leq \gamma \leq \omega_1$. When $\gamma$ approaches either side of the bound, the resolution scale is either higher or lower than desired. (a) Network structure is shown by communities whose {\it densities} are represented by each community position at the vertical axis. The node color represents ground truth communities while inter-communities edges are drawn in black dashed lines; (b) the NMI scores and the number of detected communities in relation to the resolution parameter.}
    \label{fig:gamma_error_exp}
\end{figure}

In our experiments, the extended degree-corrected planted partition model generates a network comprised of ten communities, each has ten nodes and a {\it density} $\omega_r$ illustrated by the red vertical lines in Fig.~\ref{fig:gamma_error_exp}. The background inter-community {\it density} $\omega_0$ is chosen as $0.17$, a value much smaller than any intra-community {\it densities}. The performance of community detection is measured by the normalized mutual information (NMI) metric~\cite{wagner2007comparing} that compares the detected partition with the ground truth partition used for generation. 

As Fig.~\ref{fig:gamma_error_exp}(b) shows, the generalized modularity performs well with resolution parameter in the interval $\gamma\in[1.8,4.2]$, generally matching the derived theoretical bound $\omega_0 \leq \gamma \leq \omega_1$. As $\gamma$ approaches either side of the bound, the resolution parameter is either higher or lower than desired. The asymptotic bounds defined here are derived by approximating the number of edges with the corresponding expectation in the random graph model (Eq.~(\ref{eq:approx_out2}),~(\ref{eq:approx_in2})). Hence, as $\gamma$ is getting closer and closer to either $\omega_0$ or $\omega_1$, the asymptotic results are getting further and further away from the true values, causing the NMI score to drop. This scenario is illustrated in Fig.~\ref{fig:gamma_error_exp}(a) where every community $r$ is placed at the height corresponding to its {\it density} $\omega_r$. When the resolution parameter is between the asymptotic bounds, the modularity maximization can successfully detect communities. However, when the resolution parameter is larger than the {\it density} of any of the communities, those communities are at risk of being split into smaller parts.

\begin{figure}[t!]
    \centering
    \includegraphics[width=\textwidth]{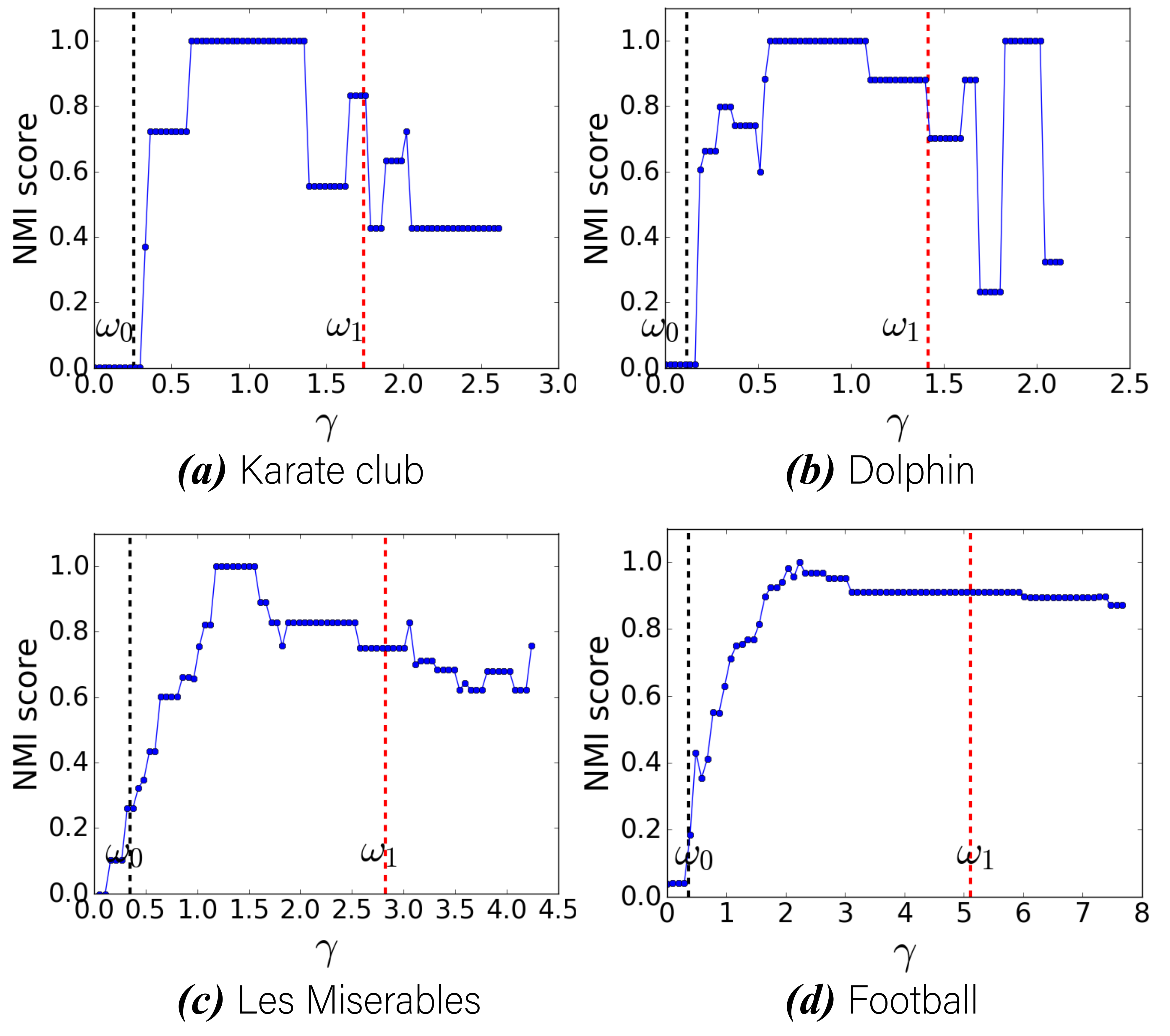}
    \caption{The alignment between the communities detected by generalized modularity maximization and the optimal $\gamma$ values listed in~\cite{newman2016equivalence}. Although the empirical networks are not generated by the extended planted partition model, maximizing the generalized modularity is optimal when the resolution parameter takes values that lie in the interval $[\omega_0, \omega_1]$. This phenomenon is also captured purely experimentally and without any theoretical justification in~\cite{fenn2009dynamic,mucha2010community,traag2013significant}.}
    \label{fig:small_networks}
\end{figure}

We test the performance of our algorithm on a range of empirical networks with $n$ nodes and $m$ edges, including the Karate club network~\cite{zachary1977information}, the dolphin social network~\cite{lusseau2003bottlenose} and the network of interactions between fictional characters in the novel {\it Les Miserables}~\cite{newman2004finding}. For each network, we compute the maximum-likelihood estimates of the background edge density $\omega_0$ and the lowest intra-community edge density $\omega_1$\footnote{See the Supplementary Material for the formal derivations of the maximum-likelihood estimates for $\omega_0$ and $\omega_1$ in the extended degree-corrected planted partition model.}, fitting an extended degree-corrected planted partition model given the number of communities $q$ and the optimal value of $\gamma$ obtained by the statistical inference~\cite{newman2016equivalence}. Then, we compute the modularity maximization results with a total of $100$ different resolution parameters in the range $[0.2,3\omega_1/2]$ and compare these results with the communities produced with $\gamma$ from this range. The subrange that produces an NMI score higher than $90\%$ is shown in Table~\ref{tab:small_networks}. As shown in Fig.~\ref{fig:small_networks}, although these empirical networks are not generated by the extended degree-corrected planted partition model, the stable intervals of resolution parameter lie inside the asymptotic lower and upper bounds established here.

\begin{table}[]
    \centering
    \caption{The maximum-likelihood estimates of $\omega_0$ and $\omega_1$ and the interval of the resolution parameter that detects communities with NMI score larger than 90\% in a range of empirical networks with $n$ nodes and $m$ edges. The number of communities $q$ used for each network is the ground truth value generally accepted in the literature. The optimal $\gamma$ values were published in~\cite{newman2016equivalence}.}
    \label{tab:small_networks}
    \begin{tabular}{ccccccc}
    \hline
    Network & $n$ & $m$ & $q$ & $\gamma$ & $(\hat{\omega}_0,\hat{\omega}_1)$ & $90\%$ interval\\
    \hline
    Karate club & 34 & 78 &  2 & 0.78 & (0.26, 1.74) & (0.63, 1.37)\\
    Dolphin social & 62 & 159 & 2 & 0.59 & (0.12, 1.42) & (0.58, 2.0)\\
    Les Miserables & 77 & 254 & 6 & 1.36 &  (0.35, 2.83) & (1.15,  1.54)\\
    \hline
    \end{tabular}
\end{table}

\subsection{Synthetic Networks}
One of the standard sources of community structures for the evaluation of community detection algorithms is the LFR benchmark~\cite{lancichinetti2008benchmark} which generates networks based on a set of pre-defined ground truth communities. In so generated networks, both the degree and community size distributions follow the power law. The main benefit of using LFR benchmark is that the ground truth communities are known. The generated networks vary with the following three parameters: $\gamma$ which is an exponent of the node degree in the power law distribution, $\beta$ which is an exponent of the community size in the power law distribution, and $\mu$ which is the {\it density} parameter that defines the fraction of all edges that have both endpoints inside the same community. The LFR benchmark is closely related to the microcanonical degree-corrected stochastic block model~\cite{peixoto2012entropy} that imposes the degree sequence as the hard constraint for all sample networks of the ensemble. Since the generated networks have the known ground truth communities, we consider the alignment of the detected communities with such ground truth a reliable quality measurement for these communities.

In our experiments, the networks generated by the LFR benchmarks have the average node degree of 9.3 and the numbers of nodes ranging from 6,000 to 15,000. The exponents $\gamma$ and $\beta$ are set to 3.0 and 1.5 respectively and the {\it density} parameter $\mu$ is equal to 0.25. We compare the multi-scale algorithm with two state-of-the-art algorithms the Louvain algorithm~\cite{blondel2008fast} and CPM algorithm~\cite{traag2011narrow}.
Since these algorithms require a single resolution parameter, we test their performance using their own resolution parameters over a wide range of commonly used values, respectively, and record the best values of NMI and ARI metrics in Table~\ref{tab:LFR1} 
The results show that the proposed multi-scale community detection algorithm performs modestly better than the two state-of-the-art modularity maximization algorithms.  It should be noted that the strongest advantage of
the multi-scale algorithm over other modularity maximization algorithms is its ability to avoid resolution limit anomaly. The results indicate that it is unlikely that the resolution limit anomaly was present in the used LFR benchmark networks.


\begin{table}[]
\centering
\caption{Performance comparison between the Louvain~\cite{blondel2008fast} algorithm, the CPM modularity maximization algorithm, and the multi-scale community detection on the LFR benchmark networks. The best results in each row are marked by bold font.}
\label{tab:LFR1}
\begin{tabular}{cc|cccc}
\hline
\multirow{2}{*}{\#Nodes} & \multirow{2}{*}{\#Edges} & \multirow{2}{*}{Metrics} & \multirow{2}{*}{Louvain} & \multirow{2}{*}{CPM} & \multirow{2}{*}{Multi-scale} \\
 &  &  &  &  &  \\
\hline
\multirow{2}{*}{30,000} & \multirow{2}{*}{139685} & ARI & 0.45 & 0.90 & {\bf 0.93}\\
& & NMI & 0.88 & 0.94 & {\bf 0.96}\\
\multirow{2}{*}{40,000} & \multirow{2}{*}{186471} & ARI & 0.69 & 0.93 & {\bf 0.95}\\
& & NMI & 0.89 & 0.96 & {\bf 0.97}\\
\multirow{2}{*}{50,000} & \multirow{2}{*}{233100} & ARI & 0.68 & {\bf 0.93} & {\bf 0.93}\\
& & NMI & 0.89 & {\bf 0.96} & {\bf 0.96}\\
\multirow{2}{*}{60,000} & \multirow{2}{*}{277207} & ARI & 0.70 & 0.89 & {\bf 0.93}\\
& & NMI & 0.87 & 0.94 & {\bf 0.95}\\
\hline
\end{tabular}
\end{table}

We compare the performance of Louvain algorithm maximizing the generalized modularity and the proposed multi-scale community detection algorithm with different values of resolution parameters on the LFR generated network with 5,000 node and 23,234 edges. As Table~\ref{tab:LFR2} illustrates, the proposed multi-scale community detection approach outperforms the Louvain algorithm. When the resolution parameter $\gamma$ takes different values, the community structures found by maximizing the generalized modularity are worse than the results produced by the multi-scale community detection algorithm. A plausible reason could be that, due to the ``plateaus'' problem mentioned above, there may not be any appropriate resolution parameter capable of avoiding all the unnecessary splits and merges at the same time. Hence, no matter what value the resolution parameter takes, maximizing the generalized modularity suffers from the resolution limit. In contrast, the proposed multi-scale community detection algorithm attempts to adapt resolution parameter to actual density in different regions of a graph. Therefore, it achieves better performance in terms of the ARI and NMI scores.

\begin{table}[]
\centering
\caption{The table compares ARI and NMI scores and execution times between the Louvain algorithm with different choice of $\gamma$ and multi-scale community detection for the LFR networks with  5,000 nodes. The column with the best results for Louvain algorithm are marked in bold font. The multi-scale algorithm starts with $\gamma_0=0.44$. Execution times are measured on a machine with the single core Intel processor.}
\label{tab:LFR2}
\begin{tabular}{c|ccccc|c}
\hline
\multirow{2}{*}{Metrics} &  \multicolumn{5}{c|}{Louvain} & {Multi-scale}\\
 & $\gamma=0.5$ & $\gamma=1.0$ & $\gamma=2$ & $\gamma=3$ & $\gamma=6$ & $\gamma_0=0.44$ \\
\hline
ARI & 0.55 & 0.76 & 0.91 & {\bf 0.93} & 0.72 & {\bf 0.94} \\
NMI & 0.77 & 0.86 & 0.92 & {\bf 0.93} & 0.88 & {\bf 0.95} \\
Time (sec) & 0.57 & 0.47 & 0.35 & {\bf 0.38} & 0.53 & {\bf 1.05} \\
\hline
\end{tabular}
\end{table}

\begin{figure}
    \centering
    \includegraphics[width=\textwidth]{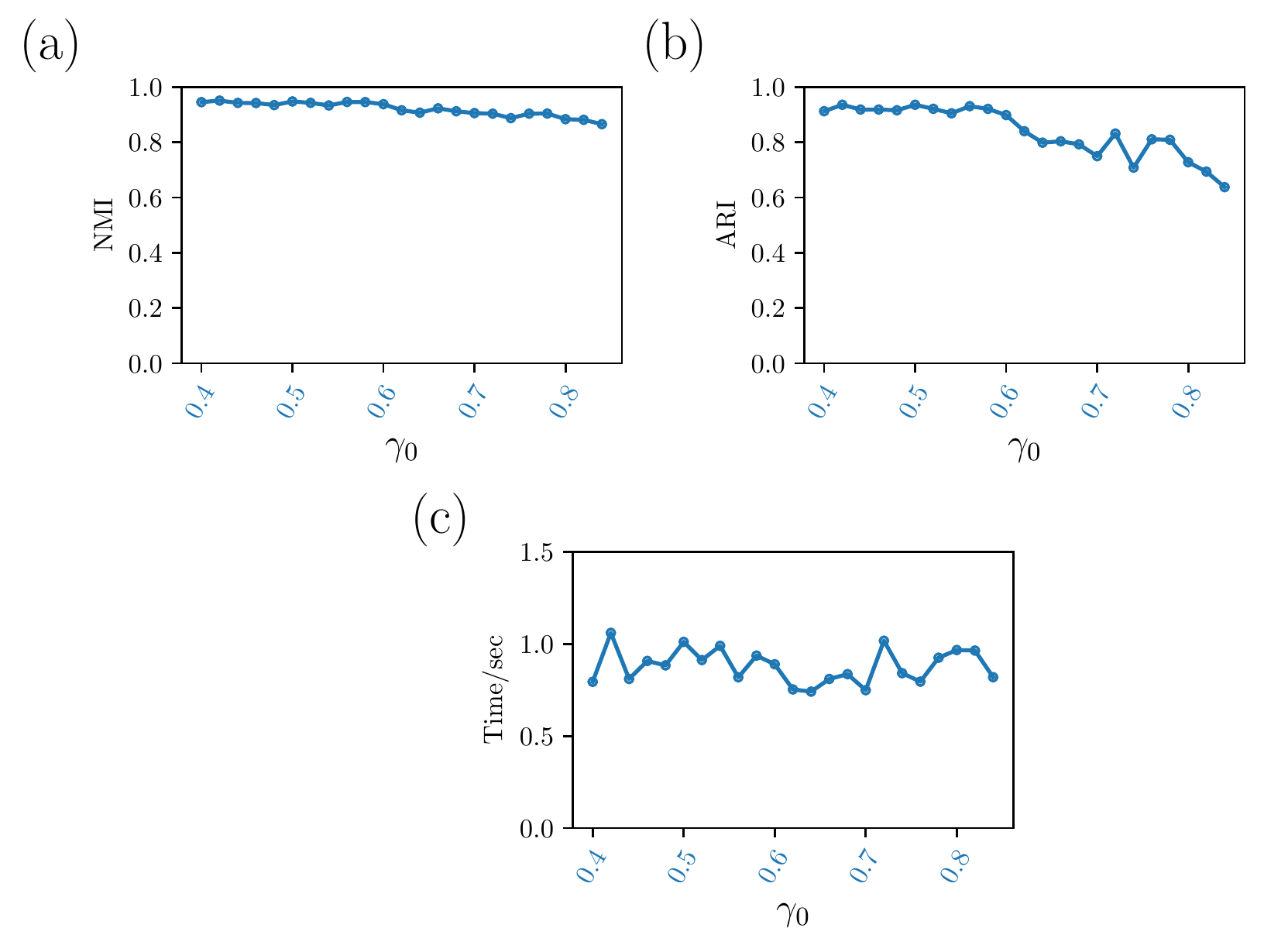}
    \caption{The plots show the sensitivity of the multi-scale community detection execution times to the value of parameter $\gamma_0$ on a LFR network with 5,000 nodes and 23,234 edges.}
    \label{fig:LFR4}
\end{figure}

Fig.~\ref{fig:LFR4} shows the multi-scale community detection algorithm's sensitivity to the choice of the initial resolution parameter $\gamma_0$. When $\gamma_0$ takes different values in the range $[0.4,0.9]$, the quality of the communities detected on a LFR network of 5,000 nodes remains relatively consistent. Using the different $\gamma_0$ values for the multi-scale algorithm, the NMI scores of the detected communities are above $0.8$. The ARI scores are also relatively stable when $\gamma_0$ is smaller than 0.7. The variation of the execution time is small as illustrated by Fig.~\ref{fig:LFR4}. 

\subsection{Real Networks}

We test the performance of the proposed multi-scale community detection algorithm on two large real networks. The first is the DBLP co-authorship network where edges connect every pair of authors who published at least one paper together. The second network is the Amazon product co-purchasing network, in which edges connect the frequently co-purchased products. According to~\cite{yang2015defining}, the publication venue, e.g., journal or conference, defines an individual ground-truth community in the DBLP network. In the Amazon network, each product category provided by Amazon defines each ground-truth community. Although these meta-data may not correlate well with the topological community structure~\cite{peel2016ground}, comparing the detected communities with the pre-defined meta-data of network provide a good hint for community detection performance in general. In addition, we only compare the detected communities with the top-5,000 communities defined in~\cite{yang2015defining} to reduce the bias introduced by the meta-data.

\begin{table}[]
\centering
\caption{The comparison of Louvain, CPM and multi-scale algorithms on real large networks using NMI and F-measure scores. The best results are marked in bold font.}
\label{tab:large}
\begin{tabular}{c|cc|cc}
\hline
 & \multicolumn{2}{c|}{Amazon} & \multicolumn{2}{c}{DBLP} \\
 &  NMI  & F-measure & NMI & F-measure\\
\hline
Louvain & 0.56 & 0.47 & 0.36 & 0.14 \\
CPM & 0.68 & 0.46 & 0.60 & 0.24 \\
Multi-scale & {\bf 0.69} & {\bf 0.49} & {\bf 0.64} & {\bf 0.26} \\
\hline
\end{tabular}
\end{table}

In Table~\ref{tab:large}, we show the NMI and F-measure for the multi-scale algorithm with $\gamma_0 = 0.5$ and CPM algorithm with resolution parameter $0.005$ on Amazon and DBLP networks. We also run Louvain algorithm with a range of resolution parameter values and show the results with the parameter value that yielded the best NMI and F-measure scores. The defintion of F-measure~\cite{wagner2007comparing} is as follows.
\begin{equation}
\textit{F-measure} = \sum_{r} \frac{|X_r|}{|V|} \max_{s \in T_{5000}} \frac{2|X_r \cap Y_s|}{|X_r|+|Y_s|}
\end{equation}
where $X_r$ is the set of nodes in the detected community $r$ while $Y_s$ is the set of nodes in the ground truth community $s$. The total number of nodes in a network is denoted as $V$ and $s \in T_{5000}$ denotes the indices of the top-5,000 high-quality ground truth communities. The results show that on DBLP data, the performance of the multi-scale algorithm exceeds Louvain performance by 86\% and CPA performance by 8\%, indicating that Louvain suffers from the effects of resolution limit anomaly much stronger than CPM algorithm does.

\begin{figure}
    \centering
    \includegraphics[width=1.0\textwidth]{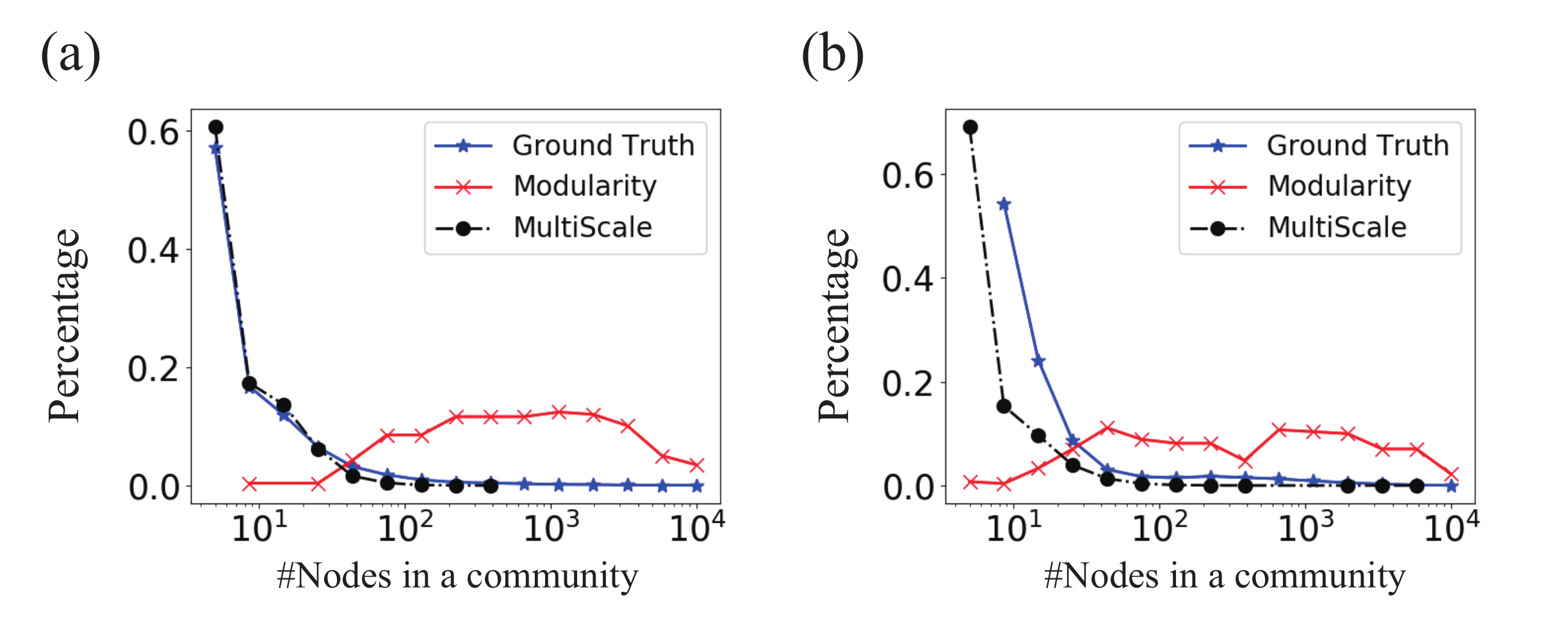}
    \caption{The performance of multi-scale, and the modularity maximization Louvain algorithms on Amazon product co-purchasing network and DBLP co-authorship network.}
    \label{fig:large}
\end{figure}

Fig.~\ref{fig:large} shows the distribution of the size of the detected communities in two real networks whose ground truth communities are defined in~\cite{yang2015defining}. The distribution of the communities detected by the proposed multi-scale algorithm matches the distribution for the so-defined ground truth communities. The modularity maximization algorithm, however, produces relatively fewer small communities due to the resolution limit. The multi-scale detection algorithm does not suffer from this issue, hence, in both the Amazon network and the DBLP network, there are many more small communities than the large ones.

\section{Conclusions} \label{sec:conclusions}
The real networks are not necessarily generated by any random graph model - all the assumptions about network generation here are only approximations of the real community structures. But we show that if the {\it densities} of edges in communities are much larger than the background {\it density} of edges across communities, the degree-corrected planted partition model is still a good approximation that avoids the resolution limit problem, when a resolution parameter is chosen from the theoretical range given in Eq.~(\ref{eq:whereisgamma}). We also show that, there is more complicated problem described as ``plateaus'' problem where no single resolution parameter exist satisfying all bounds. We propose a multi-scale community detection algorithm that requires minimal modification to the original modularity maximization method, thus preserving the high speed and robustness of modularity maximization. 

Since the degree-corrected planted partition model equivalent to the generalized modularity is much simpler than the degree-corrected stochastic block model, its performance on realistic large networks is inevitably limited. Although one can infer the block assignments of stochastic block model to obtain communities, this inference is actually much more complicated than maximizing generalized modularity. In practice, modularity maximization via the agglomerative heuristics iteratively merges neighboring blocks~\cite{clauset2004finding}. In contrast, the merging operation in the inference of stochastic block model involves many more (not necessarily adjacent) blocks as candidates~\cite{peixoto2014hierarchical}. One advantage of the inference approach is that it uses the model selection to prevent overfitting the community models~\cite{ghasemian2019evaluating}. Therefore, we believe that finding a well-defined simple quality measure that takes into account the model complexity would be of great value.

In addition, edge weighting schemes~\cite{de2013enhancing,lu2018adaptive} are effective approaches to tolerate the modularity resolution limit. The ``plateaus'' problem indicates the resolution limit arises during the heterogeneous community formation. Edge weighting may be able to fix such heterogeneity. Other future works include applying more capable models as nested hypothesis in the hypothesis testing framework and extending the multi-scale community detection algorithm to weighted and directed networks.

\section*{Acknowledgments}
This work was supported in part by the Army Research Laboratory (ARL) through the Cooperative Agreement (NS CTA) Number W911NF-09-2-0053, and by the Office of Naval Research (ONR) under Grant N00014-15-1-2640. The views and conclusions contained in this document are those of the authors and should not be interpreted as representing the official policies either expressed or implied of the Army Research Laboratory or the U.S. Government.

\section*{References}


\end{document}